**Visualizing Critical Correlations near the Metal-Insulator Transition in $Ga_{1-x}Mn_xAs$**

Anthony Richardella [1,2,*], Pedram Roushan [1,*], Shawn Mack[3], Brian Zhou[1], David A. Huse[1], David D. Awschalom[3], and Ali Yazdani [1,§]

[1]Joseph Henry Laboratories and Department of Physics, Princeton University, Princeton, New Jersey 08544, USA. [2]Department of Physics, University of Illinois at Urbana-Champaign, Urbana, IL 61801, USA. [3]Center for Spintronics and Quantum Computation, University of California, Santa Barbara, California 93106, USA.

[*] These authors contributed equally to this work.
[§] Correspondence to be addressed to: yazdani@princeton.edu

Electronic states in disordered conductors on the verge of localization are predicted to exhibit critical spatial characteristics indicative of the proximity to a metal- insulator phase transition. We have used scanning tunneling microscopy to visualize electronic states in $Ga_{1-x}Mn_xAs$ samples close to this transition. Our measurements show that doping-induced disorder produces strong spatial variations in the local tunneling conductance across a wide range of energies. Near the Fermi energy, where spectroscopic signatures of electron-electron interaction are the most prominent, the electronic states exhibit a diverging spatial correlation length. Power-law decay of the spatial correlations is accompanied by log-normal distributions of the local density of states and multifractal spatial characteristics. Our method can be used to explore critical correlations in other materials close to a quantum critical point.

Since Anderson first proposed that disorder could localize electrons in solids fifty years ago (*1*), studies of the transition between extended and localized quantum states have been at the forefront of physics (*2*). Realizations of Anderson localization occur in a wide range of physical systems from seismic waves to ultracold atomic gases, in which localization has recently been achieved using random optical lattices (*3*). In electronic systems, the signatures of localization have long been examined through electrical transport measurements (*4, 5*), and more recently using local scanning probe techniques that have imaged localized electronic states (*6, 7*). For non-interacting systems, the electronic states at the mobility edge are predicted to have a diverging localization length with scale independent power-law characteristics, which are described as being multifractal (*8*). Given the poorly understood nature of the metal-insulator transition in the presence of disorder and electron-electron interactions, direct imaging of electronic states can provide insights into the interplay between localization and interactions.

We report on scanning tunneling microscopy (STM) and spectroscopy studies of electronic states in the dilute magnetic semiconductor $Ga_{1-x}Mn_xAs$, over a range of Mn concentrations near the metal-insulator transition (x=1.5-5%). Over the last decade, $Ga_{1-x}Mn_xAs$ has emerged as a promising material for spintronic applications with a high ferromagnetic transition temperature (*9, 10*). Mn atoms substituted at Ga sites act both as acceptors that drive the metal-insulator transition and as localized spins that align at low temperatures to give rise to magnetism. The nature of the electronic states underlying magnetism in these heavily doped semiconductors is still debated. It is often assumed that the carriers that mediate magnetism in $Ga_{1-x}Mn_xAs$ are Bloch states associated with either the valence bands or extended states originating from an impurity band (*11-13*); however, the validity of these assumptions has been questioned (*14, 15*). Moreover, many recent low temperature transport studies show evidence of electron-electron interaction and weak localization of carriers even for samples with high doping levels (*16-19*). We use atomic scale imaging and high-resolution spectroscopy with the STM to visualize electronic states in $Ga_{1-x}Mn_xAs$ and to examine the spatial structure of electron-electron correlations in this system. Our results indicate that spatial heterogeneity and electronic correlations must be considered in understanding the mechanism of magnetism in highly doped semiconductors.

Fig. 1 shows STM topographs of cleaved $Ga_{1-x}Mn_xAs$ samples (2000 Å thick) grown using molecular-beam-epitaxy (MBE) on a p-type Be-doped GaAs buffer layer (*20, 21*). Prior to measurements at a temperature of 4.2 K, the degenerately doped substrates are cleaved *in situ* to expose a (110) or equivalent surface of the heterostructure in cross section. (Fig. 1A inset) The topographs show in-gap states, dominated by individual Mn acceptor wavefunctions, although other defects such as As antisites are observed as well. Using previous STM studies of samples with more dilute Mn concentrations (*22, 23*) and the results of tight-binding model calculations (*21, 24, 25*), we can identify the topographic signatures of individual Mn acceptors in layers from the surface to the 3$^{rd}$ subsurface layer (Fig. 1B). The size of an individual Mn acceptor state wavefunction is about 20 Å, due to its deep binding energy, which results in a metal-insulator transition at a relatively high level of doping (between 1-2%) in $Ga_{1-x}Mn_xAs$. Increasing the Mn concentration from weakly insulating samples with variable-range hopping resistivity at 1.5% ($T_C$=30K) (*19, 21*) to relatively conducting samples at 5% ($T_C$=86K, annealed), we find higher concentrations of dopants appear in STM topographs on top of the atomically ordered GaAs lattice (Fig. 1C). All characteristic lengths, such as dopant separation or mean free path (~10 Å), are much shorter for $Ga_{1-x}Mn_xAs$ in comparison to other semiconductors doped with shallow dopants (*15*).

Spectroscopic mapping with the STM can be used to show that Mn acceptors and other defects give rise to atomic scale fluctuations in the local electronic density of states (LDOS) over a wide range of energies. Fig. 2A shows tunneling spectra of states from within the valence band (V<0) to the conduction band edge (V>1.5), measured along a line perpendicular to the buffer layer-film interface. Contrasting the spatial dependence of electronic states in the buffer layer (with $2 \times 10^{18}$ per $cm^3$ Be acceptors) to those of $Ga_{1-x}Mn_xAs$ (with x=0.015) in Fig. 2A, we find the Mn-doped region to have strong spatial variations in the electronic states at the valence and conduction band edges and a broad distribution of states within the GaAs band gap. Increasing the Mn concentration gives rise to a larger number of in-gap states (compare Fig. 2B and 2C). These features of the local electronic structure of $Ga_{1-x}Mn_xAs$ are difficult to reconcile with a weakly disordered valence or impurity band picture, and show the importance of compensation and disorder in this compound. The Fermi energy $E_F$ lies within the range of electronic states that are spatially inhomogeneous.

In addition to strong spatial variations, electronic states of $Ga_{1-x}Mn_xAs$ are influenced by electron-electron interactions (Fig. 2D). STM spectra, spatially averaged across large areas for several samples with increasing doping levels, illustrate a strong suppression of the tunneling density of states near $E_F$. The evolution from weakly insulating (1.5%) to relatively conducting samples (5%) is well correlated with the increase in the density of states at the Fermi level, yet a suppression centered at $E_F$ is observed at all doping levels. This feature is indicative of an Altshuler-Aronov correlation gap that is expected to occur in the tunneling density of states of a disordered material on the metallic side of the phase transition (*26*), appearing as a square root singularity in the conductance near $E_F$ (Fig. 2D). Previous spectroscopic measurements of $Ga_{1-x}Mn_xAs$ using macroscopic tunneling junctions have also reported similar correlation gaps (*16*).

To determine whether there are any specific length scales associated with the spatial variation of the LDOS in $Ga_{1-x}Mn_xAs$, we now examine energy-resolved STM conductance maps. In Figure 3, we show examples of such maps at different energies relative to $E_F$ for the 1.5% doped sample. These maps show that, in addition to modulations on the length scales of individual acceptors, there are spatial structures in the local density of states with longer length scales. To characterize these variations, for each conductance map, we compute the angle averaged autocorrelation function between two points separated by r, $C(E,r)=1/(2\pi) \int d\theta \int d^2r'$ [g(E,**r'**)-$g_0$(E)]× [g(E, **r'**+**r**)-$g_0$(E)] , in which g(E,**r**) is the local value of the differential tunneling conductance that is proportional to the LDOS, $g_0$(E) is the average value of the conductance at each energy E. Displaying C(E,r) in Fig. 4A, we find a dramatic increase of the long-distance correlations near $E_F$. At this energy, the correlations remain measureable to length scales well beyond that of single Mn acceptor states, which dominate the behavior on short length scales at all energies. The increased correlation length can also be seen directly in the size of the patches of high and low conductance (Fig. 3C). We have observed the enhanced spatial correlations near $E_F$ for all doping levels examined in this study (up to 5%, Fig. 4B); however this effect is most dramatic for the least doped samples (1.5%) closest to the metal-insulator transition. Control experiments on Zn or Be doped GaAs samples show no evidence of any special length scale or of a sharp eak near the $E_F$ in the autocorrelation function.

Continuous phase transitions, such as the metal-insulator transition, are typically characterized by a correlation length, which describes the exponential decay of spatial fluctuations when a system is tuned near the phase transition. At the critical point, this correlation length diverges and spatial fluctuations and other physical properties display power-law spatial characteristics. In the non-interacting limit, the transition between a metal and an insulator occurs by tuning the chemical potential relative to the mobility edge. Mapping the spatial structure of the electronic states as function of energy can be used to determine the correlation length and to probe the critical properties for such a transition between extended and localized states (*4, 5, 8*). In our experiments, the distance dependence of the energy-resolved autocorrelation function C(E,r) for the 1.5% sample (Fig. 4C) appears to follow a power-law at $E_F$, while at nearby energies it falls off exponentially (see inset). These observations, together with the apparent divergence of the correlation at a specific energy, are indeed signatures of the critical phenomena associated with a metal-insulator transition. However, our observation that the longest-ranged correlations are centered at $E_F$, as opposed to some other energy, which could be identified as a mobility edge, signifies the importance of electron-electron interactions in the observed correlations.

Given the importance of electron-electron interactions, the conductance maps are perhaps more precisely identified as probing the spatial nature of quasiparticle excitations of a many-body system rather than simply imaging single-electron states in the non-interacting limit. Currently there are no theoretical models of the real space structure of these excitations near the metal-insulator transition in a strongly interacting and disordered system, although there is continued effort in understanding the nature of such transitions in the presence of interactions (*5, 27*). Nevertheless, we suspect that the correlation length associated with these excitations will become shorter due to multi-particle processes and inelastic effects at energies away from $E_F$. Our experimental results for the least conducting sample (1.5%) indicate that the correlation length ξ is indeed suppressed away from $E_F$, roughly following $(E-E_F)^{-1}$ (dashed line in Fig. 4A). At $E_F$, for this sample, these correlations decay in space following a power-law $r^{-\eta}$, with η=1.2 ± 0.3.

Despite the importance of strong interactions, many of the predictions for the non-interacting limit still appear to apply. Weakly disordered extended states are expected to show Gaussian distributions of the LDOS indicating that these states have a finite probability to be present over the entire system. In contrast, near the metal-insulator transition wide distributions are expected, especially in local quantities such as the LDOS, which begin to cross over from Gaussian to lognormal distributions even in the limit of weak localization (*28, 29*). Spectroscopic maps of the density of states at $E_F$ for three different dopings (Fig. 5A-C) show different degrees of spatial variations; however, their histograms (Fig. 5D) are similar in being skewed log-normal distributions where the mean is not representative of the distribution due to rare large values. Decreasing the doping skews the distribution further in a systematic fashion away from Gaussian and toward a log-normal distribution. For comparison, a histogram of the LDOS for states deep in the valence band for the least doped sample (dark gray circles in Fig. 5D) shows a Gaussian distribution.

Based on the predictions for the non-interacting limit, we expect critical states to have a spatial structure that is multifractal in nature. This property is directly related to the scale invariant nature of critical wavefunctions and has been examined in great detail by numerical simulations of the single-particle quantum states near an Anderson transition (*8*). Multifractal patterns, which are ubiquitous in nature, are usually described by analysis of their self-similarity at different length scales through their singularity spectrum f(α). Physically, f(α) describes all the fractal dimensions embedded in a spatial pattern, such as those associated with a quantum wavefunction and its probability distribution. It is calculated by splitting the probability distribution into sets of locations {$r_i$} that share a common exponent α, where the distribution scales locally with distance like $|\Psi(r_i)|^2 \sim L^{-\alpha}$, and measuring the fractal dimension of each set (*8, 21*). A variety of techniques have been developed to compute f(α), which has been used to distinguish between various models of the Anderson transition (*21, 30*). Application of such an analysis to our conductance maps (Fig. 5D, inset), shows an f(α) spectrum that is peaked at a value away from two, which is indicative of anomalous scaling in a two-dimensional map. The f(α) spectrum also shows a systematic shift with decreasing doping, indicating a trend from weak toward strong multifractality with decreasing doping. In contrast, these signatures of multifractal behavior are absent for states deep in the valence band (gray curve) that, despite the strong disorder, show scaling consistent with those expected for extended states.

Our findings suggest that proximity to the metal-insulator transition and electronic correlations may play a more significant role in the underlying mechanism of magnetism of $Ga_{1-x}Mn_xAs$ than previously anticipated. Beyond its application to understand the nature of states $Ga_{1-x}Mn_xAs$, our experimental approach provides a direct method to examine critical correlations for other material systems near a quantum phase transition. In principle, experiments at the lowest temperatures for samples closest to the metal-insulator transition should provide accurate measurements of power-law characteristics that can be directly compared to theoretically predicted critical exponents.

## Figure Captions

**Figure 1.** STM topography of the in-gap states of GaMnAs. (**A**) STM topograph (+1.5 V, 20 pA) of $Ga_{0.985}Mn_{0.015}As$ over a 1000 Å by 500 Å area. The inset shows a topograph (+1.8 V, 20 pA) of the heterostructure junction between the Mn-doped and Be-doped layers of GaAs. The size of this area is 80 Å by 125 Å. (**B**) A topograph (+1.5 V, 30 pA) over a smaller area of size 150 Å by 150 Å of $Ga_{0.985}Mn_{0.015}As$. Several substitutional Mn's in various layers are identified in the lower part of the panel and marked by which layer they are in, with the surface labeled as zero. An As anti-site is also shown. (**C**) STM topograph (+1.5 V, 20 pA) of the $Ga_{0.95}Mn_{0.05}As$ sample after annealing for 6 hours at 200 °C. The size of the area is 150 Å by 195 Å. (B) and (C) have the same scale.

**Figure 2.** Local dI/dV spectroscopy for various dopings. (**A**) The differential conductance (dI/dV) ($\Delta V=10$ mV) across a line normal to the growth surface of length 1000 Å. The first 200 Å is the Be-doped GaAs buffer layer, and 800 Å of 1.5% Mn-doped GaAs follows. The junction between the two is marked by a dashed line. (**B**), (**C**) The dI/dV spectra ($\Delta V=5$ mV) for energies close to Fermi level, across a line of length 450 Å over the 1.5% sample (B), and the 5% sample (C). Top of the valence band and the in-gap states are shown. (**D**) The spatially averaged differential conductance for several samples. The inset shows the same data as the main panel, with the square root of the voltage on the horizontal axis.

**Figure 3.** Spectroscopic Mapping in $Ga_{0.985}Mn_{0.015}As$. The differential conductance (dI/dV) measurements over an area of 500 Å by 500 Å show the spatial variations in the LDOS for the states in the valence band as well as states deep inside the semiconductor gap: (**A**), -100 mV; (**B**), -50 mV; (**C**), 0mV; (**D**), +50 mV; (**E**), +100 mV; and (**F**), +150 mV.

**Figure 4.** Correlation length for Mn dopings close to the metal-insulator transition. The autocorrelation was calculated from LDOS maps and is presented on a logarithmic scale for (**A**) $Ga_{0.985}Mn_{0.015}As$ and (**B**) as-grown $Ga_{0.95}Mn_{0.05}As$. The dashed line is ~ $(E-E_F)^{-1}$ and is a guide to the eye. (**C**) The autocorrelation for the dI/dV map at the Fermi level of $Ga_{0.985}Mn_{0.015}As$, as well as one valence band (-50 mV) and one in-gap energy (+50 mV) are plotted. The inset shows the autocorrelation for the same energies on a semi-logarithmic scale.

**Figure 5.** The spatial variations of the LDOS at the Fermi level, their histogram and multifractal spectrum. The LDOS mapping of a 700 Å by 700 Å area of (**A**) $Ga_{0.985}Mn_{0.015}As$, (**B**) $Ga_{0.97}Mn_{0.03}As$, and (**C**) $Ga_{0.95}Mn_{0.05}As$. (**D**) The normalized histogram of the maps presented in parts (A) to (C). The local values of the dI/dV are normalized by the average value of each map. The inset shows the multifractal spectrum, $f(\alpha)$, near the value $\alpha_0$ where the maximum value occurs. For comparison, the results of similar analysis over a LDOS map at -100 mV (valence band states) for the 1.5% doped sample are also shown.

# References

1. P. W. Anderson, *Phys. Rev.* **109**, 1492 (1958).
2. A. Lagendijk, B. van Tiggelen, D. S. Wiersma, *Physics Today* **62**, 24 (2009).
3. A. Aspect, M. Inguscio, *Physics Today* **62**, 30 (2009).
4. P. A. Lee, T. V. Ramakrishnan, *Rev. Mod. Phys.* **57**, 287 (1985).
5. D. Belitz, T. R. Kirkpatrick, *Rev. Mod. Phys.* **66**, 261 (1994).
6. S. Ilani *et al.*, *Nature* **427**, 328 (2004).
7. K. Hashimoto *et al.*, *Phys. Rev. Lett.* **101**, 256802 (2008).
8. F. Evers, A. D. Mirlin, *Rev. Mod. Phys.* **80**, 1355 (2008).
9. H. Ohno *et al.*, *Nature* **408**, 944 (2000).
10. D. Chiba *et al.*, *Nature* **455**, 515 (2008).
11. T. Dietl, H. Ohno, F. Matsukura, J. Cibert, D. Ferrand, *Science* **287**, 1019 (2000).
12. K. S. Burch *et al.*, *Phys. Rev. Lett.* **97**, 087208 (2006).
13. T. Jungwirth *et al.*, *Phys. Rev. B* **76**, 125206 (2007).
14. T. Dietl, *J. Phys. Soc. Jpn.* **77**, 031005 (2008).
15. C. P. Moca *et al.*, *Phys. Rev. Lett.* **102**, 137203 (2009).
16. S. H. Chun, S. J. Potashnik, K. C. Ku, P. Schiffer, N. Samarth, *Phys. Rev. B* **66**, 100408(R) (2002).
17. D. Neumaier *et al.*, *Phys. Rev. Lett.* **99**, 116803 (2007).
18. L. P. Rokhinson *et al.*, *Phys. Rev. B* **76**, 161201 (2007).
19. B. L. Sheu *et al.*, *Phys. Rev. Lett.* **99**, 227205 (2007).
20. R. C. Myers *et al.*, *Phys. Rev. B* **74**, 155203 (2006).
21. Materials and Methods are available as supporting material on Science Online.
22. D. Kitchen, A. Richardella, J. M. Tang, M. E. Flatte, A. Yazdani, *Nature* **442**, 436 (2006).
23. J. K. Garleff *et al.*, *Phys. Rev. B* **78**, 075313 (2008).
24. J. M. Tang, M. E. Flatte, *Phys. Rev. Lett.* **92**, 047201 (2004).
25. T. O. Strandberg, C. M. Canali, A. H. MacDonald, *Phys. Rev. B* **80**, 024425 (2009).
26. B. L. Altshuler, A. G. Aronov, *Solid State Comm.* **30**, 115 (1979).
27. D. M. Basko, I. L. Aleiner, B. L. Altshuler, *Annals of Physics* **321**, 1126 (2006).
28. I. V. Lerner, *Physics Letters A* **133**, 253 (1988).
29. B. L. Altshuler, V. E. Kravtsov, I. V. Lerner, *Physics Letters A* **134**, 488 (1989).
30. A. Chhabra, R. V. Jensen, *Phys. Rev. Lett.* **62**, 1327 (1989).
31. This work was supported by grants from ONR, ARO, NSF, and the NSF-MRSEC program through the Princeton Center for Complex material. P.R. acknowledges a NSF graduate fellowship.

**Supporting Online Material**
www.sciencemag.org
Materials and Methods
Figs. S1, S2, S3, S4

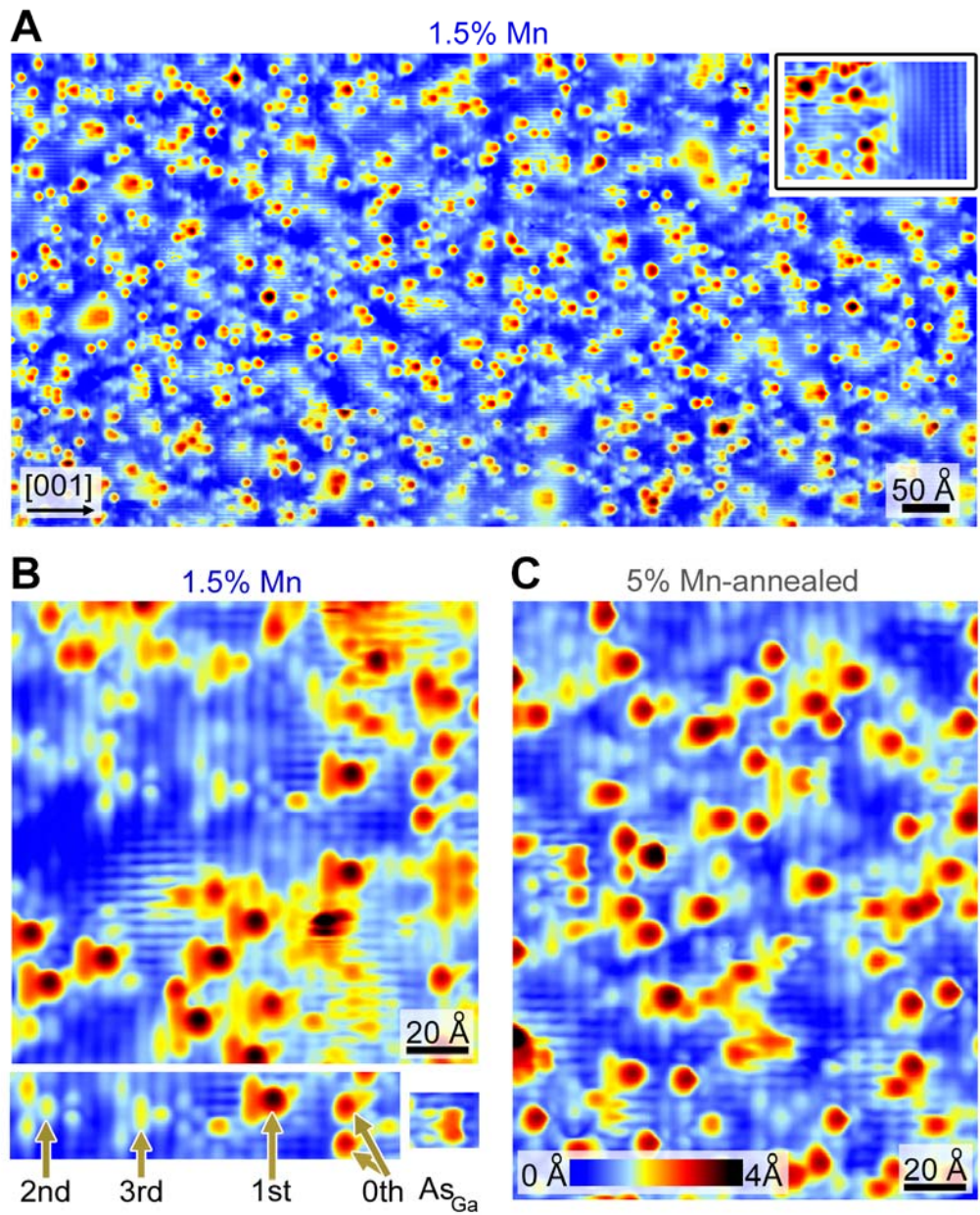

Figure 1

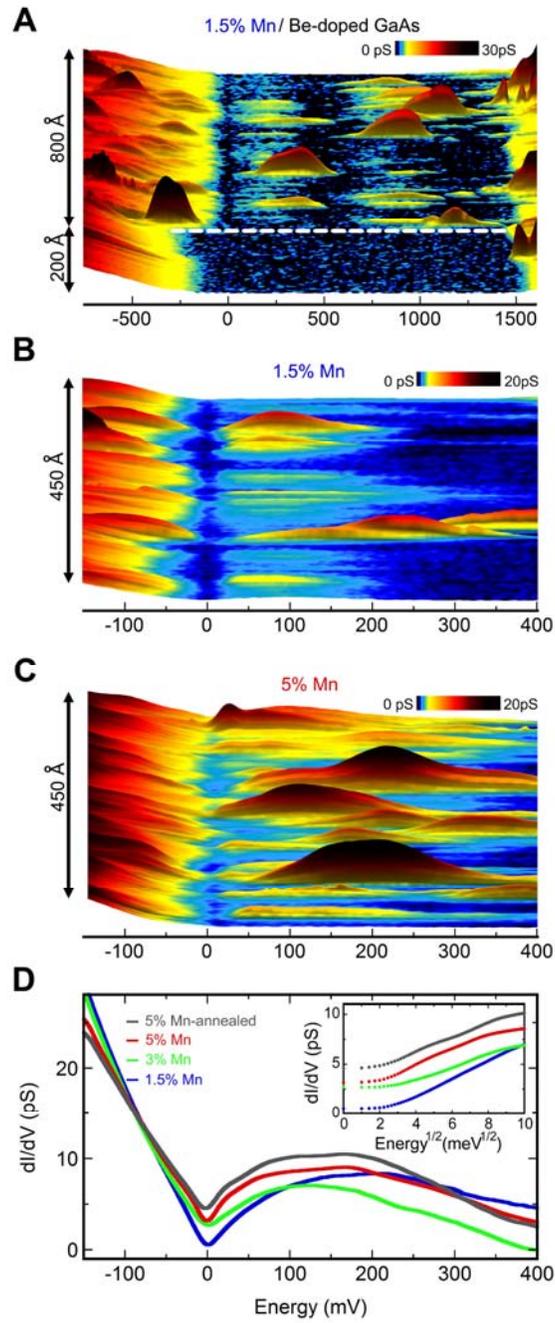

Figure 2

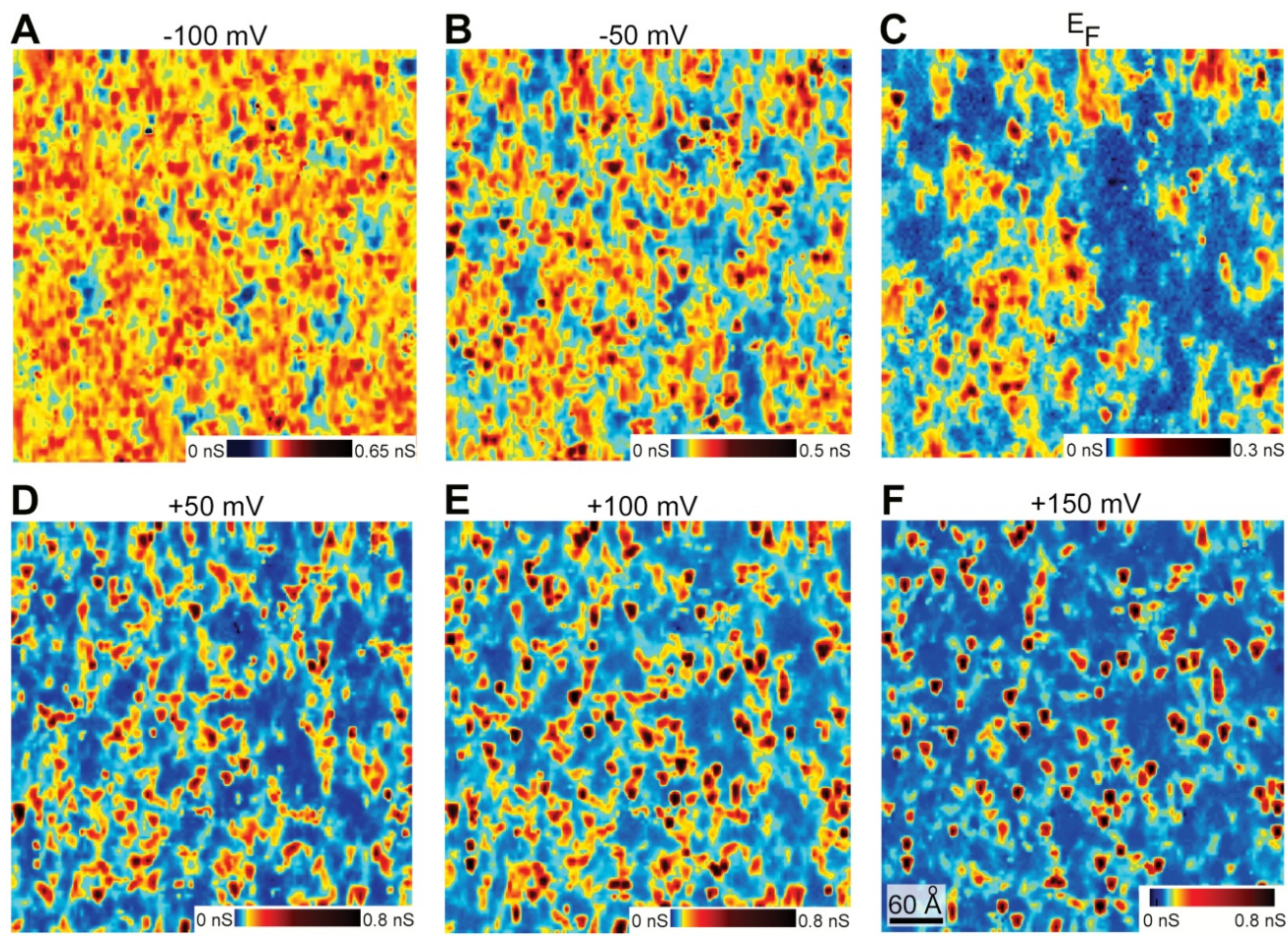

Figure 3

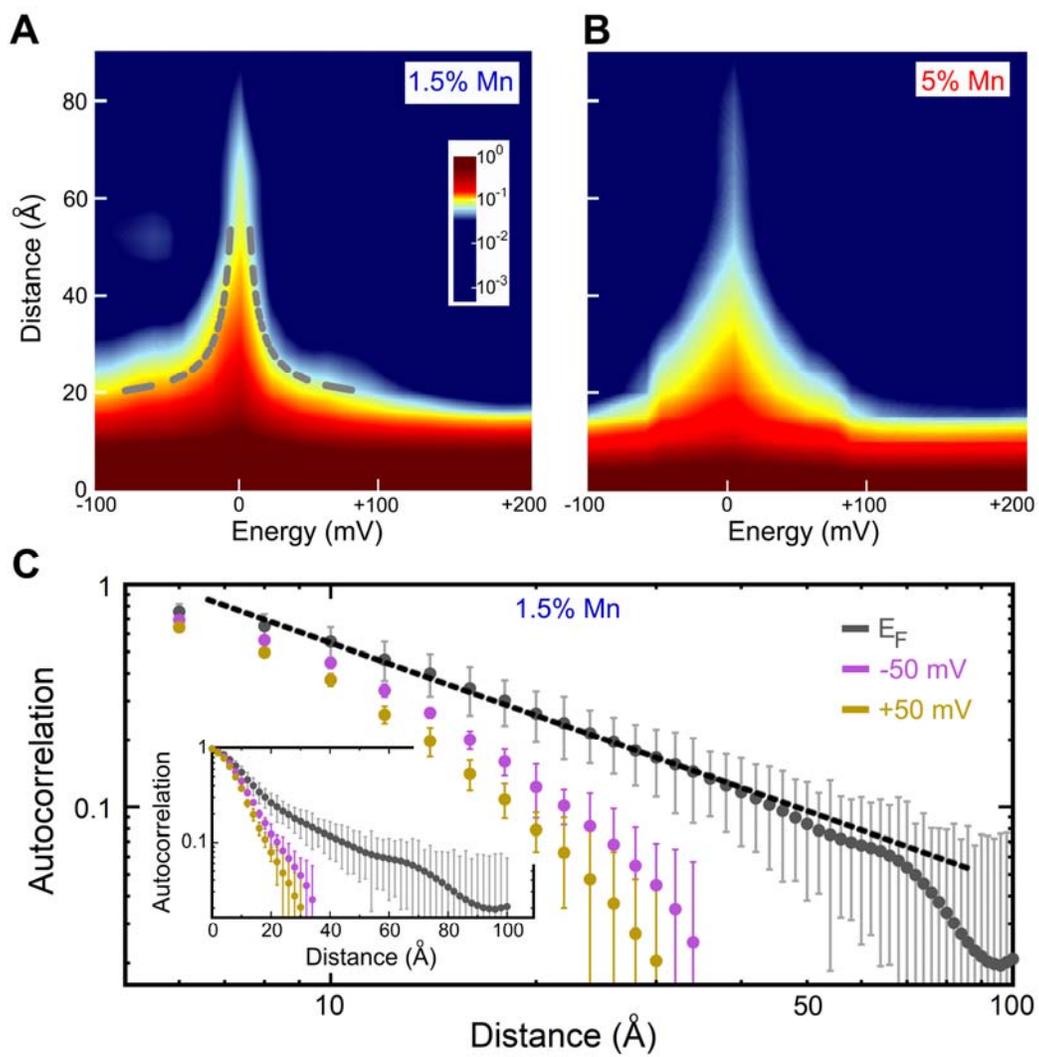

Figure 4

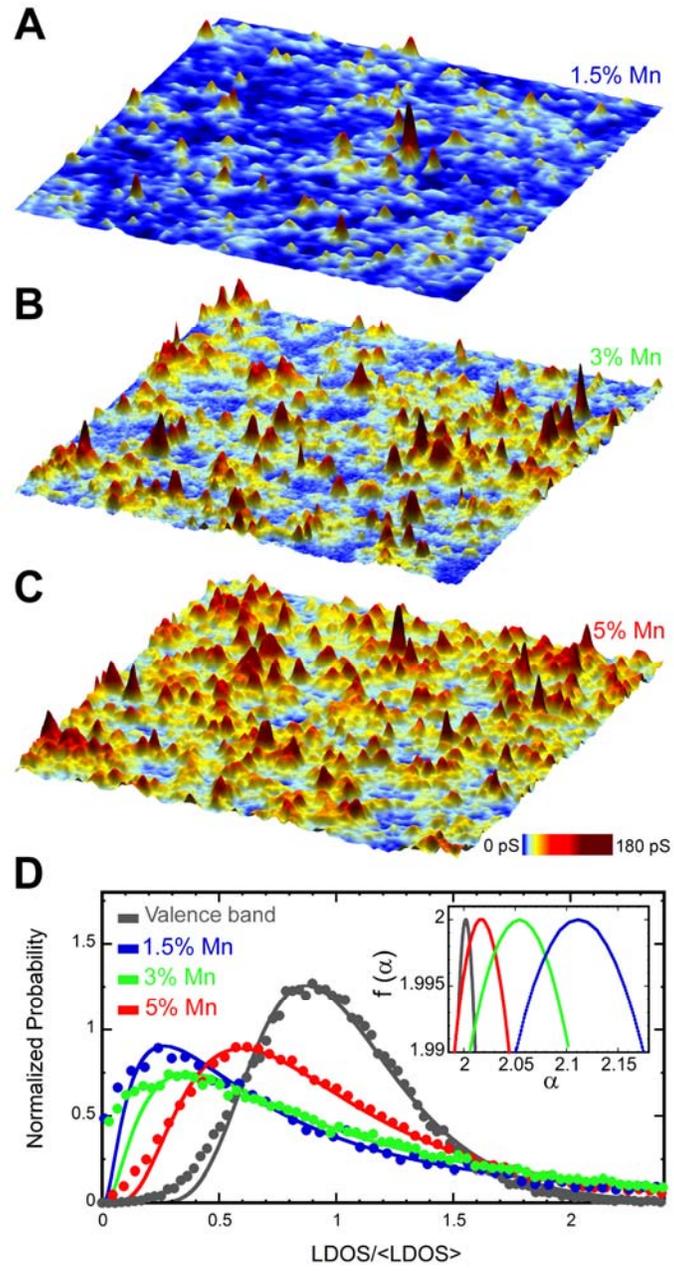

Figure 5

Supporting Online Material for

**Visualizing Critical Correlations near the Metal-Insulator Transition in Ga$_{1-x}$Mn$_x$As**

Anthony Richardella [1,2,†*], Pedram Roushan [1*], Shawn Mack[3], Brian Zhou[1], David A. Huse[1], David D. Awschalom[3], and Ali Yazdani [1§]

[1]Joseph Henry Laboratories and Department of Physics, Princeton University, Princeton, New Jersey 08544, USA.
[2]Department of Physics, University of Illinois at Urbana-Champaign, Urbana, IL 61801, USA.
[3]Center for Spintronics and Quantum Computation, University of California, Santa Barbara, California 93106, USA.

* These authors contributed equally to this work.
† Present address: Dept. of Physics, The Pennsylvania State University, University Park, PA 16802
§ Correspondence to be addressed to: yazdani@princeton.edu

## Materials and Methods

**Ga$_{1-x}$Mn$_x$As Samples**

All samples studied were grown in a Varian Gen-II MBE system manufactured by Applied Epi. Samples with 1.5%, 3% and 5% Mn concentrations, which were grown for this experiment, consisted of a 500 μm thick, 2 inch diameter, degenerately doped n-type GaAs substrate followed by a 150 nm thick MBE grown 1x10$^{18}$ cm$^{-3}$ silicon doped n-type buffer, a 150 nm thick MBE grown 2x10$^{18}$ cm$^{-3}$ beryllium doped p-type buffer and finally 200 nm of Ga$_{1-x}$Mn$_x$As. During growth, the substrate temperature was monitored via band-edge thermometry, and the temperature stability during growths was typically ±5 °C. The growth rate of GaAs at the center of the wafer was 0.7 ML/ s as calibrated by RHEED intensity oscillations of the specular spot at 580 °C. The samples were grown in a "non-rotated" geometry to produce a spatial gradient of arsenic flux. We characterized different sections of the wafer and performed the STM experiment on the sections closest to stoichiometry. The ferromagnetic transition temperatures of the as-grown samples studied were 30K, 43K and 70 K for 1.5%, 3% and 5% Mn-doped samples, respectively. A 5% sample annealed for 5 hours at 200 °C with a T$_C$ of 86 K was also studied.

Resistivity trends provide complementary evidence for electron-electron interactions. We measured sheet resistance in a magneto-cryostat via the van der Pauw technique with soldered-indium contacts at the sample corners. Samples were cleaved from wafer regions near stoichiometry. Parallel conduction through the Be-doped p-GaAs and Ga$_{1-x}$Mn$_x$As layers is likely; however, the magnitude of resistance indicates little conduction through the conductive n-GaAs substrates, and we observe resistivity characteristics common to $Ga_{1-x}Mn_xAs$ films on semi-insulating substrates. In particular, the enhanced resistivity below 8 K (Fig. S1) indicates Altshuler-Aronov electron-electron interactions (*S1, S2, and references in main text*).

**Details of STM Procedures**

The samples were thinned to less than 100μm in thickness to facilitate the creation of atomically flat surfaces at the heterostructure during the cleaving process. Larger thicknesses tended to result either in a disordered surface or a myriad of closely spaced atomic steps. The cleaves were performed *in-situ* under ultra high vacuum to expose either a (1 1 0) or (1 1̄ 0) surface and the sample was then inserted into the cold cryostat, where the STM head resides, within a few minutes. A tip made of iridium was prepared by field emission on a clean metal surface and checked for consistent spectroscopic signatures before the study of each sample. Conducting substrates had to be used to allow the STM tip to tunnel into the body of the sample before finding the heterostructure at the front edge. The tip was then moved in small steps towards the heterostructure until it was correctly positioned. Once there, the tip could also be moved along the length of the $Ga_{1-x}Mn_xAs$ and various areas of the surface were studied, all of which had similar properties within a single sample.

**Details of Data Analysis**

*Single Impurity Identification:*

Mn dopants were identified by comparison to previous STM studies of isolated Mn acceptors and theoretical calculations (Fig. S2), which show a strong dependence of the observed wavefunction on the distance the Mn resides from the surface (as referenced in the text). This, together with the alternation of the wavefunction being centered on or between the observed atomic rows of the surface depending on whether the Mn resides in an even or odd subsurface layer, allows the determination of the depth of the dopant (*S3*). A similar procedure can be followed for the identification of As antisites that are present in the material due to its low temperature growth. As antisites have been observed in low temperature grown GaAs by STM and have been well reproduced by *ab initio* calculations (*S4, S5*).

For each concentration several samples were studied, and the probed areas on each sample were several millimeters apart. Our STM's field of view is about 5 microns, allowing us to observe across the entire Mn-doped layer. The topographies presented in Fig. 1 of the report are typical of what we have observed across various regions of the samples studied. No topographic or spectroscopic signatures of MnAs nano-islands where ever observed, nor were tendencies for the Mn to cluster seen, indicating that all samples studied had homogeneous compositions. The topographies like the one shown in Fig. 1C for 5%-annealed samples show only a higher concentration of Mn-dopants.

*Search for Signatures of Mn Clustering:*

Analysis of whether there is any tendency for Mn to cluster is difficult at high concentrations due to overlap of the wavefunctions, which makes it difficult to identify individual Mn dopants. However, the unique topographic signature of Mn in the surface layer at energies deep in the valence band states does allow us to identify all substitutional Mn in the 0th layer (*S6*). To provide a quantitative analysis, we present (Fig. S3) the cumulative probability of random pair separation; i.e., for a given value of *r*, *P(<r)* is the probability of two randomly chosen Mn dopants being less than a distance *r* from one another. For homogeneously distributed dopants, this probability scales with the area ($r^2$), and the dopants fill the field of view, leaving no void area. The measured probability shows a power law behavior with a power very close to 2 for an extended range of distances (10 Å < r < 1000 Å), indicating the absence of clustering in this compound. In contrast to a single exponent observed here, the tendency toward cluster formation would result in different exponents at different length scales, starting with values higher than 2 in small scales and smaller than 2 for large scales. Unfortunately, such topographies deep in the valence band over large areas were not made of the 5% samples. However, we have no reason to believe they would behave differently.

*Multifractal Analysis:*

The multifractal spectrum, f(α), is calculated from the generalized inverse participation numbers,

$$P_q = \int d^d r |\psi(r)|^{2q} \propto L^{-\tau_q} \tag{S2}$$

which correspond to different moments of the spatial probability distribution of the wave function. In general, the different moments scale with different powers of the length, L, which can be the size of a box within which the probability density of the wavefunction is measured. The multifractal spectrum is the Legendre transform of $\tau_q$,

$$f(\alpha) = \alpha q - \tau(q),$$
$$\alpha = \frac{d\tau(q)}{dq} \tag{S3}$$

The robust method for calculating the *f(α)* curve proposed by Chhabra and Jensen was used,

$$f(q) = \lim_{L \to 0} \frac{\sum_i \mu_i(q,L) \log[\mu_i(q,L)]}{\log L}$$
$$\alpha(q) = \lim_{L \to 0} \frac{\sum_i \mu_i(q,L) \log[P_i(L)]}{\log L} \tag{S4}$$

and the results were consistent with curves generated directly from τ(q)(*S7*).

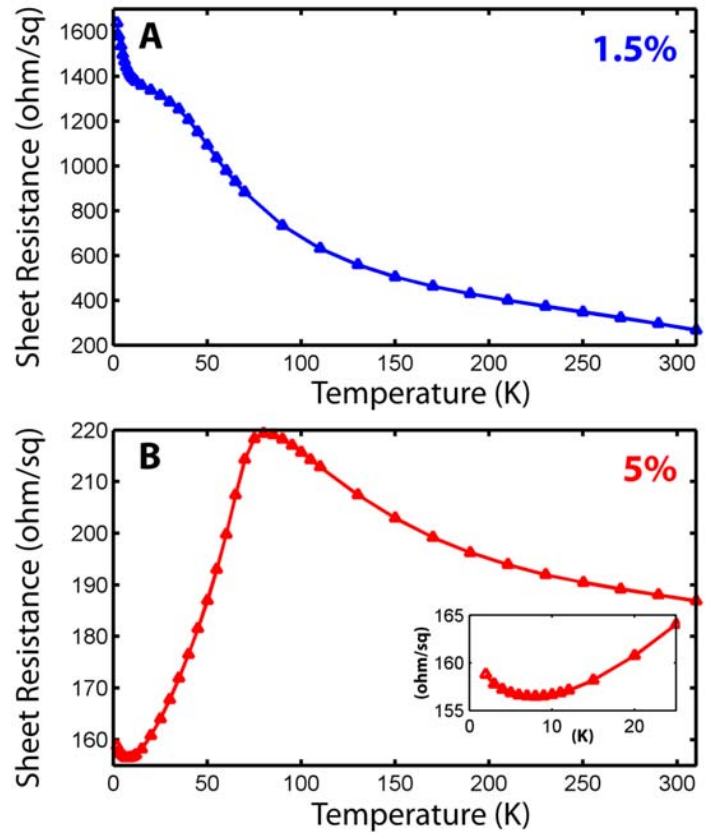

**Fig. S1.** Sheet resistances vs temperature. (**A**) 1.5% sample shows insulating-like behavior with decreasing temperature. (**B**) 5% curve shows metallic resistance with the typically observed resistance hump around $T_C$ and an uptick in resistance at the lowest temperatures due to correlations (inset).

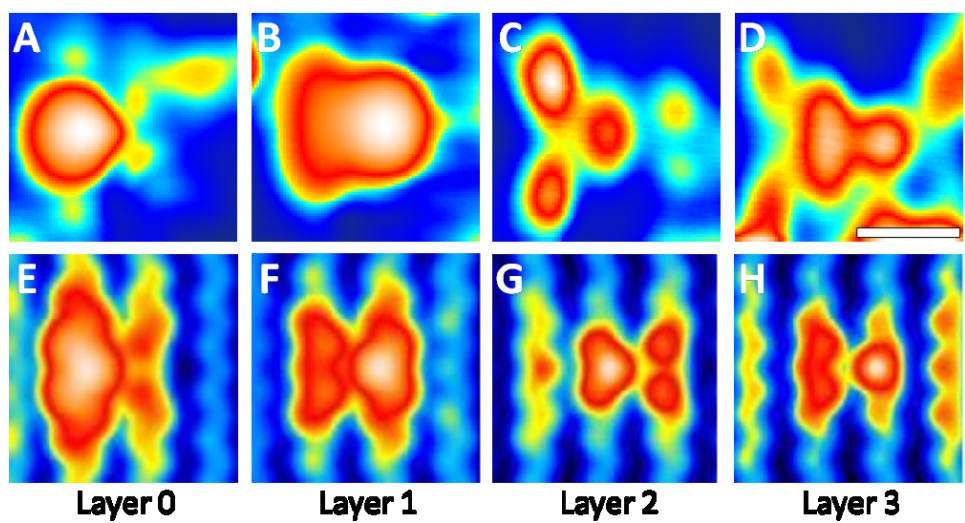

**Fig. S2.** Comparison of STM topographs of Mn dopants in various layers beneath the (110) surface with simulated STM images from tight binding calculations. (**A-D**) STM topographs (1.5V, 20pA) of Mn dopants located in surface layer (Layer 0) down to three layers beneath the surface, respectively. The scale bar is 10Å. (**E-H**) Simulated STM images from Ref. S4 of tight binding calculations of the Mn acceptor state in bulk GaAs as viewed from the plane of the dopant and up to three layers away.

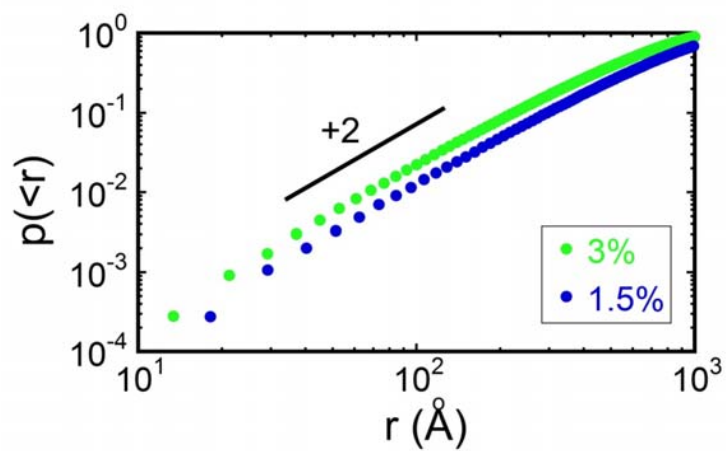

**Fig. S3.** Study of the possibility of clustering by calculating the probability of correlation between Mn pairs in different locations. For a randomly chosen pair of Mn, *P(<r)* gives the probability of the pair having distance less than *r*. The calculated probability scales very close to $r^2$ (solid line) for an extended range of distances, demonstrating the uniform distribution of Mn dopants for 1.5% and 3% doped samples.

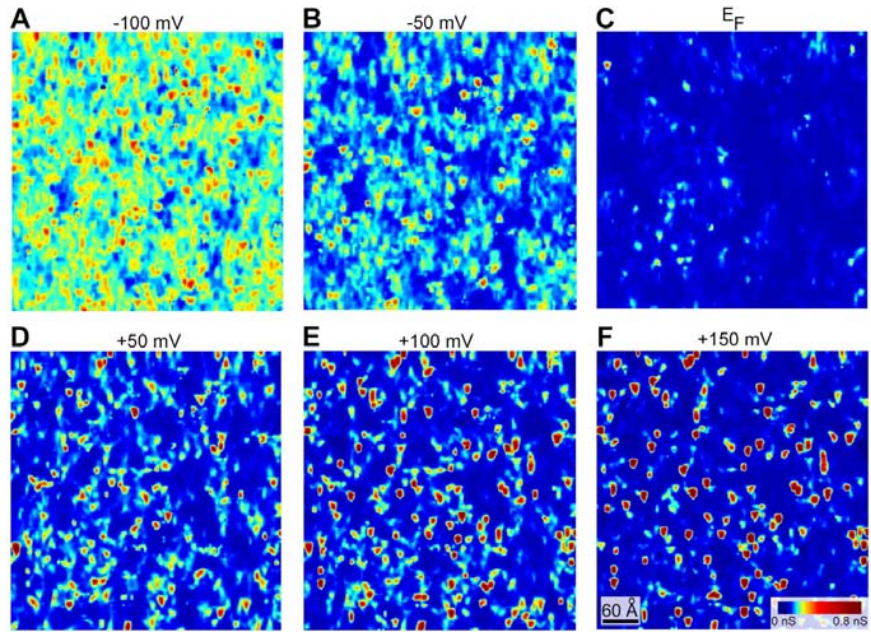

**Fig. S4.** Spectroscopic mapping in $Ga_{0.985}Mn_{0.015}As$ using the same colorscale for all panels. This obscures fine details in the individual maps but facilitates direct comparisons between states at different energies. The differential conductance (dI/dV) measurements over an area of 500 Å by 500 Å, show the spatial variations in the LDOS for the states in the valence band as well as states deep inside the semiconductor gap: (**A**), -100 mV; (**B**), -50 mV; (**C**), 0mV; (**D**), +50 mV; (**E**), +100 mV; and (**F**), +150 mV.

# References

S1. L. P. Rokhinson *et al.*, *Phys. Rev. B* **76**, 161201 (2007).

S2. D. Neumaier *et al.*, *Phys. Rev. Lett.* **103**, 087203 (2009).

S3. R. de Kort, M. C. M. M. van der Wielen, A. J. A. van Roij, W. Kets, H. van Kempen, *Physical Review B* **63**, 125336 (2001).

S4. R. M. Feenstra, J. M. Woodall, G. D. Pettit, *Physical Review Letters* **71**, 1176 (1993).

S5. R. B. Capaz, K. Cho, J. D. Joannopoulos, *Physical Review Letters* **75**, 1811 (1995).

S6. D. Kitchen, A. Richardella, J. M. Tang, M. E. Flatte, A. Yazdani, *Nature* **442**, 436 (2006).

S7. A. Chhabra, R. V. Jensen, *Physical Review Letters* **62**, 1327 (1989).